\documentclass[pra,aps,twocolumn,notitlepage,superscriptaddress,showpacs,nofootinbib]{revtex4-2}

\usepackage{enumerate,appendix}
\usepackage{amsmath, amsthm, amssymb,commath}
\usepackage{color,calc,graphicx}
\usepackage[usenames,dvipsnames,svgnames,table,cmyk,hyperref]{xcolor}
\usepackage[colorlinks]{hyperref}
\hypersetup{
	colorlinks = true,
	urlcolor = {blue},
	citecolor = {magenta},
	linkcolor= {blue}
}

\usepackage{graphicx}
\usepackage{amsmath}
\usepackage{latexsym}
\usepackage{bbm}
\newcommand{\steve}[1]{{\color{black} #1}}

\newcommand{\ket}[1]{|#1\rangle}
\newcommand{\bra}[1]{\langle#1|}
\newcommand{\braket}[2]{\langle#1|#2\rangle}
\newcommand{\bracket}[3]{\langle#1|#2|#3\rangle}
\newcommand{\ketbra}[2]{|#1\rangle\langle#2|}

\newcommand{\tr}{{\rm tr}}

\usepackage[charter,cal=cmcal,sfscaled=false]{mathdesign}
\usepackage{booktabs}
\usepackage{multirow}
\usepackage{subfigure}
\usepackage{dcolumn}
\usepackage{mathrsfs}
\usepackage{diagbox}
 \usepackage{array}
 \usepackage{makecell}

\begin{document}

\title{Certification of a Nonprojective Qudit Measurement using Multiport Beamsplitters}


\author{Daniel Mart\'inez}
\author{Esteban S.~G\'{o}mez}
\affiliation{Departamento de F\'{\i}sica, Universidad de Concepci\'on, 160-C Concepci\'on, Chile}
\affiliation{Millennium Institute for Research in Optics, Universidad de Concepci\'on, 160-C Concepci\'on, Chile}

\author{Jaime Cari\~ne}
\affiliation{Millennium Institute for Research in Optics, Universidad de Concepci\'on, 160-C Concepci\'on, Chile}
\affiliation{Departamento de Ingenier\'ia El\'ectrica, Universidad Cat\'olica de la Sant\'isima Concepci\'on, Concepci\'on, Chile.}

\author{Luciano Pereira}
\affiliation{Instituto de F\'{\i}sica Fundamental IFF-CSIC, Calle Serrano 113b, Madrid 28006, Spain}
\author{Aldo Delgado}
\author{Stephen P.~Walborn}
\affiliation{Departamento de F\'{\i}sica, Universidad de Concepci\'on, 160-C Concepci\'on, Chile}
\affiliation{Millennium Institute for Research in Optics, Universidad de Concepci\'on, 160-C Concepci\'on, Chile}

\author{Armin Tavakoli}
\affiliation{Institute for Quantum Optics and Quantum Information - IQOQI Vienna, Austrian Academy of Sciences, Boltzmanngasse 3, 1090 Vienna, Austria}
\affiliation{Institute for Atomic and Subatomic Physics, Vienna University of Technology, 1020 Vienna, Austria}

\author{Gustavo Lima}
\affiliation{Departamento de F\'{\i}sica, Universidad de Concepci\'on, 160-C Concepci\'on, Chile}
\affiliation{Millennium Institute for Research in Optics, Universidad de Concepci\'on, 160-C Concepci\'on, Chile}

\begin{abstract}
Generalised quantum measurements go beyond the textbook concept of a projection onto an orthonormal basis in Hilbert space. They are not only of fundamental relevance but have also an important role in quantum information tasks. However, it is highly demanding to certify that a generalised measurement is \steve{indeed required to explain} the results of a quantum experiment in which only the degrees of freedom are assumed to be known. Here, we use state-of-the-art multicore optical fiber technology to build multiport beamsplitters and faithfully implement a \steve{seven-outcome generalised measurement in a four-dimensional Hilbert space} with a fidelity of $99.7\%$. We apply it to perform an elementary quantum communication task and demonstrate a success rate that cannot be simulated in any conceivable quantum protocol based on standard projective measurements on quantum messages of the same dimension. Our approach, \steve{which} is compatible with modern photonic platforms, showcases an avenue for faithful and high-quality implementation of genuinely nonprojective quantum measurements beyond qubit systems.
\end{abstract}

\maketitle

\textit{Introduction.---}
The traditional concept of a quantum measurement is to project a state onto a complete set of orthogonal projectors in Hilbert space. However, the modern formulation of quantum theory allows for more general, qualitatively different, measurements \cite{NielsenChuang}. These generalised, or nonprojective, measurements enrich quantum theory as they \steve{often} cannot be simulated with standard, projective, measurements  \cite{DAriano2005, Oszmaniec2017}. Moreover, generalised measurements are \steve{important and useful} in quantum information science, including tasks such as  quantum state discrimination \cite{Barnett2009, Bae2015, Tavakoli2021b}, state tomography \cite{Derka1998, Renes2004}, quantum key distribution \cite{Bennett1992, Englert2004, Bengtsson2020}, quantum random number generation \cite{Acin2016, Tavakoli2021}, entanglement detection \cite{Shang2018, Bae2019} and self-testing \cite{Tavakoli2020, Steinberg2021}. This has led to considerable experimental interest in nonprojective measurements (see e.g.~\cite{Clarke2001, Mosley2006, Du2006, Durt2008,  Medendorp2011, Waldherr2012, Pimenta2013, Agnew2014, Zhao2015, Bent2015, Bian2015, Schiavon2016, Sosa2017, Bouchard2018, Huang2021, Cai2021}). Given their conceptual appeal and their role as a quantum information resource, it is natural to investigate methods for realising such measurements and to certify that they are indispensable, as compared to standard measurements, to optimally perform concrete tasks.

Generalised measurements are meaningful in Hilbert spaces of fixed dimension, i.e. when the experimental degrees of freedom are known, due to the possibility of Neumark dilations \cite{NielsenChuang}. Therefore, a forceful framework to certify that an experiment is based on a generalised measurement is to consider all involved quantum operations as uncharacterised, up to their Hilbert space dimension \cite{Tavakoli2018}. Recent experiments have demonstrated certification of generalised measurements on a qubit \cite{Gomez2016, Gomez2018, Smania2020, Tavakoli2020}. However, given also the increasing relevance of higher-dimensional quantum information (see e.g.~\cite{Erhard2018, Friis2019, Erhard2020}), it is natural go beyond the simplest quantum system and thus consider generalised measurements of higher dimension.

Faithfully implementing certifiably nonprojective measurements on a quantum system dimension $d$ is challenging. Firstly, a proper implementation requires one to enable the read-out of all the possible outcomes in every round. This entails the introduction of ancillary degrees of freedom with which the system is to be entangled. The \steve{complexity of the setup} is expected to scale with $d$; already qutrit systems would be considerably more demanding than their qubit counterparts \cite{Tabia2012}. Secondly, certifying a generalised measurement in such a minimalistic scenario, i.e. when only the dimension is known and devices can be classically correlated, requires highly precise experimental apparatuses. For instance, the smallest known total visibility required to certify a qubit generalised measurements is $97.0\%$ \cite{Acin2016, Gomez2016, Mironowicz2019, Tavakoli2020, Smania2020}, which in known schemes becomes even higher when $d>2$ \cite{Rosset2019}.

Here, we \steve{demonstrate a new path towards realising high-quality generalised quantum measurements on $d$-dimensional  systems by using modern space-division multiplexing optical fiber technology \cite{Richardson_natphoton_2013}.} Specifically, we show that multiport beamsplitters, built within new multicore optical fibers (MCFs) \cite{Ming,Optica_2020}, can be used for implementing \steve{a seven outcome measurement in a $d=4$ dimensional Hilbert space} that cannot be simulated with projective measurements. To enable a semi-device-independent certification \steve{of this}, we introduce a quantum communication task that involves an uncharacterised sender that emits four-dimensional states and an uncharacterised receiver that performs measurements to access information encoded by the sender. We prove that an optimal quantum implementation requires the use of generalised measurements. \steve{Due to the high quality of the device, we achieved an average visibility greater than $99.7\%$, which allowed us to experimentally certify \steve{(by more than three standard deviations)} that our measurement is nonprojective} by achieving a success rate that surpasses the best success rate possible in any quantum protocol based on projective measurements.  Our approach expands the reach of state-of-the-art photonic devices to faithful implementation of higher-dimensional nonprojective measurements at nearly perfect fidelity.  Moreover, it is fully compatible with integrated photonics \cite{Ding_2017, Politi_2008, Obrien2015, FabioTritter, Thompson2018} and can, therefore, find applications in different quantum information scenarios.
\par
\steve{
\textit{Generalized measurements with multiport beam splitters.---} 
A multiport beam splitter (MBS) is a device described by unitary matrix $U_D$, which couples $D$ input optical modes to $D$ output modes. When used as a measurement device, they can be used to implement generalized measurements on a $d$-dimensional quantum system. The case $D=d$  corresponds to projective measurement. For $d < D$, a finite number of different rank-1 Positive Operator Value Measurements (POVMs) can be realized, as illustrated in Fig. \ref{povmmbs}. This is achieved using a sub-set of $d$ input modes as the principal system and the remaining $D-d$ modes as an ancilla system. Thereby, the total number of different possible POVMs is given by the number of ways to choose $d$ out of $D$ elements without repetition, which is ${D!}/{d!(D-d)!}$.  Taking in to account that the phase on each input mode of the system can be modulated, the rank-1 POVM elements are given by $\Pi_j=\ket{\eta_j}\bra{\eta_j}$, with
\begin{align}
\ket{\eta_j}=\Phi^\dagger_{k_1\cdots k_{d}} M_{k_1\cdots k_{d}}\ket{j},
\end{align}
where $k_1\cdots k_{d}$ are indices corresponding to the input modes defining the quantum system, $\Phi_{k_1\cdots k_{d}}$ is a diagonal matrix containing the relative phases, and $M_{k_1\cdots k_{d}}$ is the portion of the matrix $U_D^\dagger$ restricted to the system space. In this way, a number of generalized measurements can be implemented on a $d$ dimensional photonic quantum system by connecting the optical modes in different ways.  Examples for $D=7$, $d=4$ can be found in the appendix. As we will show in our experiment, the use of new fiber-based MBS devices allows for implementation of generalized measurements of sufficient quality to allow for certification.  
\begin{figure}
	\includegraphics[width=7cm]{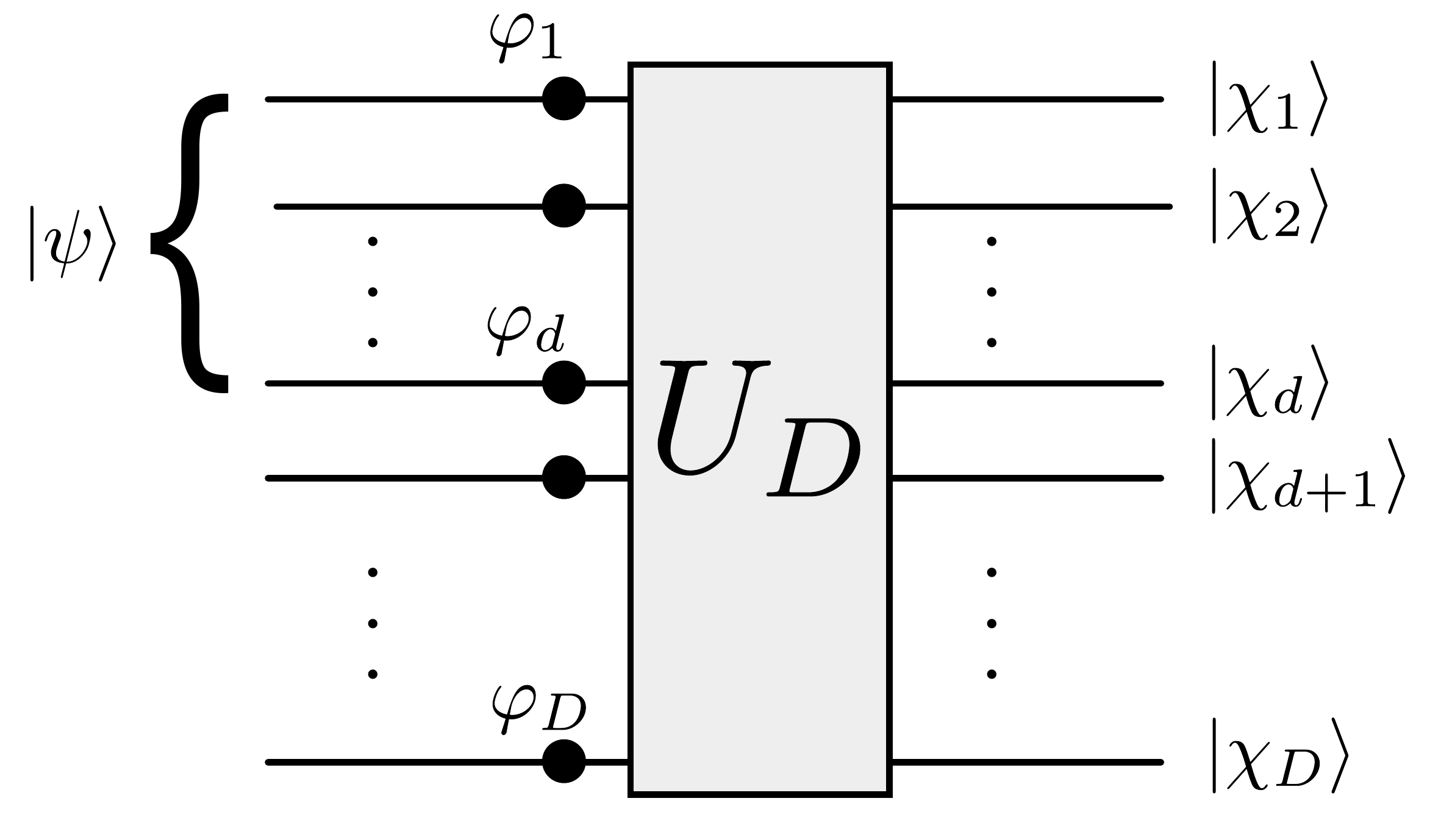}
	\caption{Generalized measurments scheme using a multiport beam splitter. The horizontal lines represent the optical modes, the black circles represent phase shifts $\varphi_j$, the box represents the multi-core beam splitter. We encode the input quantum state only in the first $d$ modes. Each output mode is associated with a POVM element.}
	\label{povmmbs}
\end{figure}
}
\begin{figure}
	\centering
	\includegraphics[width=\columnwidth]{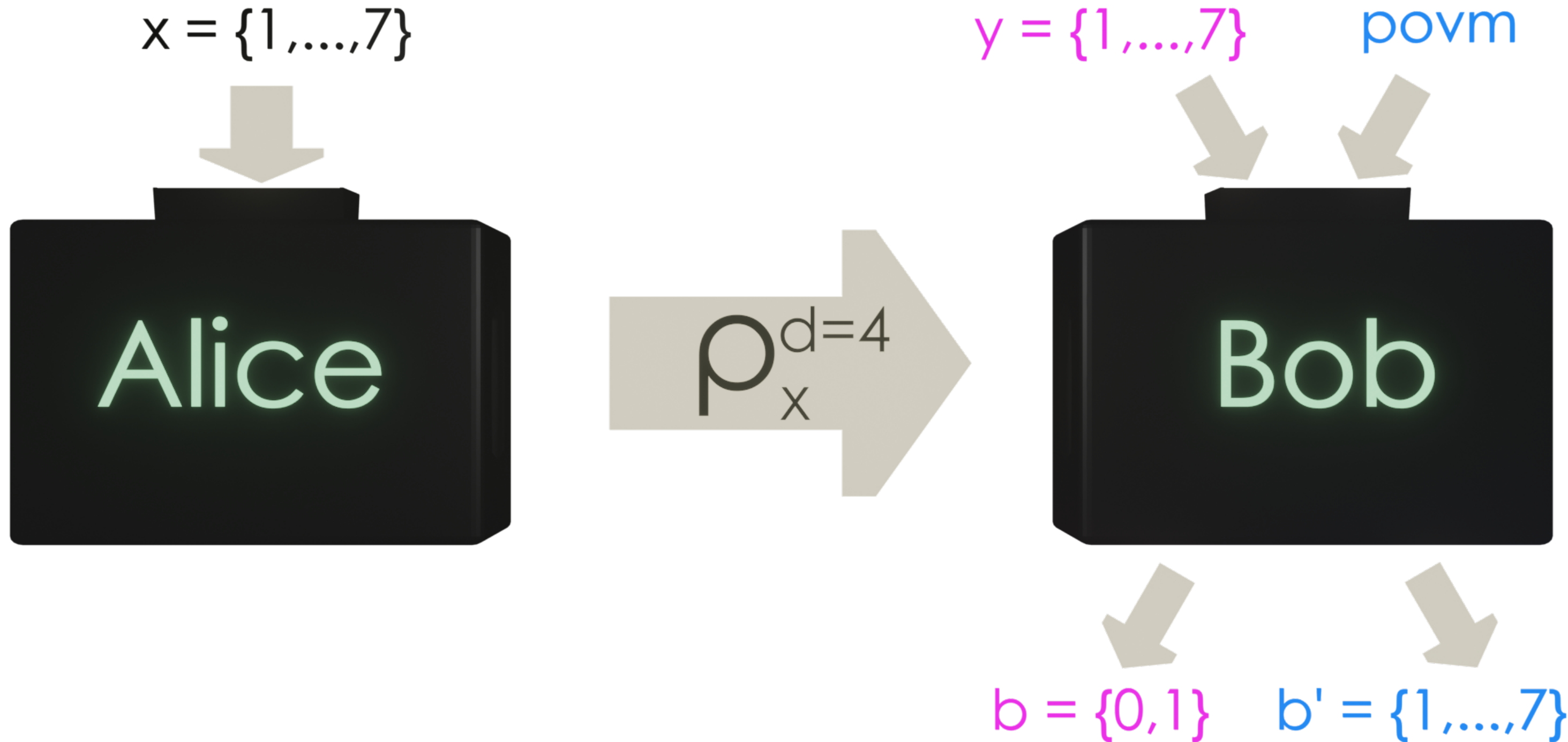}
	\caption{Communication scenario for certifying that the setting $\mathbf{povm}$ corresponds to a nonprojective four-dimensional quantum measurement.}\label{scenario210110}
\end{figure}

\textit{Certification scheme.---}  Consider the scenario illustrated in Fig~\ref{scenario210110}. A sender, Alice, selects one of seven possible classical inputs $x\in\{1,\ldots,7\}$ and encodes it into a four-dimensional quantum state, $\rho_x$, that is sent to Bob. Bob chooses between two actions. Either, he selects some $y\in\{1,\ldots,7\}$ and performs a dichotomic quantum measurement $\{E_{b|y}\}$ with the goal of learning whether  $x=y$ ($b=1$) or $x\neq y$ ($b=0$). A correct answer in the former (latter) is rewarded with two (one) points. Alternatively, Bob can select an eighth input, labelled $\mathbf{povm}$, corresponding to a seven-outcome measurement $\{M_{b'}\}$ for $b'\in\{1,\ldots,7\}$, with goal of learning Alice's input ($b'=x$). A correct answer is rewarded with three points.
From the Born rule, the total number of points in the communication task is
\begin{equation}
W=\sum_{x,y=1}^7 \left(1+\delta_{x,y}\right)\tr(\rho_x E_{\delta_{x,y}|y})+3\sum_{x=1}^7 \tr(\rho_x M_x).
\label{W}
\end{equation} 
Moreover, Alice and Bob are allowed to stochastically synchronise their quantum operations via a shared source of classical randomness. However, due to the linearity of the scoring function, the optimal strategy is deterministic.

We have determined the largest value of $W$ (labelled $W_\text{proj}$) achievable in a quantum protocol where Bob's measurements are projective and four-dimensional. To this end, we first note that a deterministic projective measurement \steve{in this scenario} can have at most four outcomes. An upper bound on $W_\text{proj}$ is then obtained via the method of semidefinite programming relaxations of \steve{fixed}-dimensional quantum correlations \cite{Navascues2015}. See \steve{appendix} for details. To obtain a tight bound, we have additionally exploited the symmetries of $W$ \cite{Rosset2019}, which \steve{can} dramatically boost the efficiency of such computations (examples in \cite{Rosset2019, Aguilar2018, Martinez2018, Pauwels2021}). We have proved tightness by saturating (up to solver precision) the upper bound with a lower bound  obtained from an explicit quantum model found via alternating convex search in the communication scenario (see e.g.~\cite{Feix2015, Tavakoli2017}). We thus find that projective quantum models are limited by
\begin{equation}\label{proj}
W_\text{proj}=62.5152.
\end{equation}

To certify that the input $\mathbf{povm}$ corresponds to a generalised measurement, we must show that there exists a quantum protocol that outperforms the bound \eqref{proj}. Using the above methods, one can show that the largest quantum value is $W_\text{quant}=62.75$. However, we focus on a protocol based on a generalised measurement that is partularly suitable for implementation in our photonics platform. Choosing $\mathbf{povm}$ as this measurement, we have numerically found  states $\{\rho_x\}$ and measurements $\{E_{b|y}\}$ (see  Appendix~\ref{AppStrategy}) that achieve the desired certification, specifically reaching $W_\text{prtcl}=62.6982>W_\text{proj}$.

\textit{Experiment---} Our experiment is based on modern multicore optical fibers created for space-division multiplexing in classical telecommunication, which is a technique that uses multiple spatial optical modes for increasing data communication capacity \cite{Richardson_natphoton_2013}. These fibers are composed of \steve{several 8$ \mu$m diameter cores  that reside within the same cladding material}. The core separation ensures that the light coupling between them is greatly reduced (<40dB/km), such that each core works as an independent and isolated transmission channel. Multicore fiber technology was recently introduced as a toolbox for high-dimensional quantum information processing \cite{GuixReview_2020}. The basic idea is to build multi-arm interferometers within a multicore fiber, taking advantage of the low loss, intrinsic stability and high-quality optical mode provided by the fiber, to implement high-fidelity unitary operations in higher-dimensional quantum systems. These interferometers have been proven to be useful not only for new fundamental investigations \cite{Lee17, Lee19, Mate21, Esteban21}, but also for quantum communication systems \cite{Glima17, Bacco17, Lio20}, quantum randomness generation \cite{Optica_2020}, and quantum computing \cite{Taddei_2020}. Yet, all these works have been limited to a fixed core geometry and, therefore, capable of only implementing standard projective measurements. Motivated by the potential of generalized measurements for quantum information processing, we now show how multicore fibers of different core geometries can be combined together for implementing high-fidelity general quantum measurements in higher-dimensions. To our knowledge, this is the first time that general measurements are implemented in a dimension $d>2$ and with a fidelity that allows for their semi-device independent certification.

\begin{figure*}[t]
\centering
\includegraphics[width=0.9\linewidth]{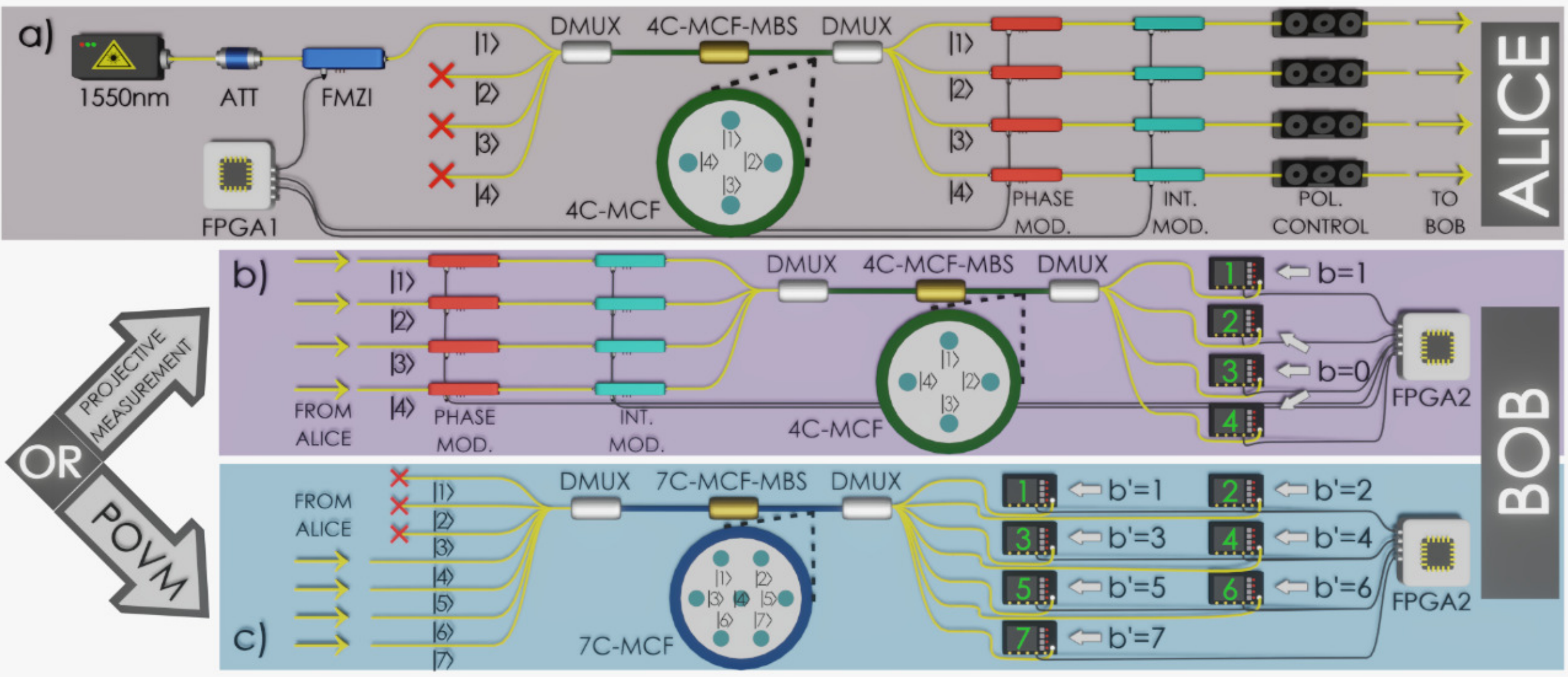}
\caption{Experimental setup for certification of generalized measurements. The combined preparation (Alice) and measurment (Bob) stages form a four-path interferometer.  a) Using four-core fiber beam splitter and additional fiber devices along with phase modulators and intensity modulators controlled by FPGA1, Alice can prepare arbitrary four-dimensional ququarts encoded in single photons. b) For projective measurements, Bob uses a scheme similar to Alice's preparation stage, allowing for projection onto a four-dimensional basis that he can define by modulating the phase and intensity of the four fiber modes. c) In the \textbf{povm} measurement, Bob connects the four input cores from Alice to a seven-core fiber beam splitter.  The additional optical modes serve as an ancilla system.  The seven-outcome measurement is nonprojective.  Single photon counts are registered by FPGA2.  See main text for additional details.}
\label{Fig_setup}
\end{figure*}

The experimental setup is illustrated in Fig.~\ref{Fig_setup}. It is essentially a four-path interferometer consisting of two main blocks: the preparation stage (Alice) and the measurement stage (Bob), who can implement projective or povm measurements. To prepare the states considered in the protocol, we use as the light source a continuous-wave telecom laser operating at 1546 nm [see Fig.~\ref{Fig_setup}(a)]. It is connected to a fiber-pigtailed amplitude modulator (FMZ), which \steve{generates 5 ns long} pulses at a repetition rate of 2 MHz. Attenuators are then used to set the average number of photons per pulse to $\mu=0.2$, such that our source can be seen as a good approximation of a non-deterministic source of single photons \cite{GisinQC2002}. In this configuration, the probability of having pulses containing at least one photon is $P( n\geq 1|\mu=0.2)\approx 18\%$ and $90.3\%$ of the non-null pulses contain only one photon. The photons are sent through \steve{a single-mode fiber (SMF) connected to a} built-in fiber \steve{demultiplexer (DMUX)}, connecting four independent SMFs to one core of a four-core MCF \cite{MCF_Demux1}. With this, light can be sent from a standard single-mode fiber into one core of a MCF or vice versa. Here, only one core of the 4C-MCF is illuminated. The DMUX is then connected to a 4C-MCF based multiport beamsplitter (MBS)  \cite{Ming,Optica_2020} that coherently distributes the signal over the four fiber cores which define a four dimensional quantum state of a single photon.  \steve{Due to the symmetry of the 4C-MCF structure, there is a} close to ideal 25$\%$ coupling between all four cores of the fiber. The 4CF-MBS corresponds to a unitary Hadamard matrix described by \cite{Optica_2020}
\begin{equation}U_{4}=\frac{1}{2}
\begin{bmatrix}
 1 & 1 & 1 & 1 \\
 1 & 1 & -1 & -1 \\
 1 & -1 & 1 & -1 \\
 1 & -1 &  -1 & 1 \label{Matrix_Theo}
\end{bmatrix}.\end{equation} Thus, the 4CF-MBS takes a photon in the logical core state $\ket{j} (j=1,...4)$ to an equally-weighted four-dimensional ``ququart" of the form $\ket{\psi_j}=\frac{1}{2} \sum_{k=1}^4u_{kj}\ket{k}$, where $u_{kj}=\pm 1$ are the entries of the matrix \eqref{Matrix_Theo}. Likewise, it can be used to map superposition states into logical basis states: $U_{4} \ket{\psi_j}=\ket{j}$. In the experiment, the 4CF-MBS has a mean fidelity with Eq.~(\ref{Matrix_Theo}) of $F= 0.995\pm0.003$ \cite{Optica_2020}. The output defines four paths of a Mach-Zehnder interferometer which will be used to prepare the states and implement the measurements required to violate the bound of Eq.~\eqref{proj} in the protocol described above. This is done in the same spirit of phase-coding quantum cryptography where the states are prepared by Alice controlling the initial part of the interferometer, while Bob performs the measurements by applying a second set of phases at the final part of it \cite{Bennett1992}. In order to control the initial quantum state entering the interferometer, a second DMUX is used to couple the 4CF-MBS to four SMFs, so that each path can be sent through a  phase (PM) and amplitude (IM) fiber-pigtailed modulator, as depicted in Fig.~\ref{Fig_setup}(a). The general ququart state that Alice prepares is then given by $| \chi  \rangle = \frac{1}{\sqrt{N}}\sum^{4}_{j=1} \alpha_j e^{i \phi^A_j}  | j  \rangle,$ where $\alpha_j$ and $\phi^{A}_j$ are the transmissivity and relative phase of core $j$, and $N$ is a normalization constant. The seven required states in the protocol are prepared by applying specific voltage values to these modulators.

To implement the required projective and generalized measurements, Bob can resort to two different measurement procedures schematically shown in Fig.~\ref{Fig_setup}(b) and Fig.~\ref{Fig_setup}(c), respectively. \steve{The projective measurement bases are implemented using a system that is similar to Alice's preparation stage. A second set of intensity and phase modulators are used to adjust the amplitude and phase, and are then connected to another 4CF-MBS. This thus closes a 4-arm 4-output interferometer [see Fig.~\ref{Fig_setup}(b)]. Bob can thus project onto a basis defined by states $| \zeta_j  \rangle = \frac{1}{\sqrt{N}}\sum^{4}_{k=1} u_{kj} \beta_k e^{i \phi^A_k}  |k  \rangle,$ where $\beta_k$ and $\phi^{B}_k$ are the transmissivities and relative phases defined by Bob, and $N$ is a normalization constant.}  


To detect the photons, the outputs of the 4CF-MBS are connected through a DMUX and SMFs to four avalanche photo-detectors (APD). They are commercial InGaAs single-photon detection modules that triggered by the laser modulation signal, and configured with 5 ns detection gate and $10 \%$ detection efficiency. In reference to the certification scheme, the outcome $b=1$ corresponds to the APD in output mode 1, while the outcome $b=0$ statistics is given in terms of the sum of APDs counts in outputs 2, 3 and 4.

The seven-outcome povm measurement adopted in the protocol can be realised using a 7$\times$7 MBS built in a 7C-MCF \cite{Optica_2020}. The core geometry of the 7C-MCF is shown schematically in the inset of Fig.~\ref{Fig_setup}(c). \steve{Even though the different core geometries of the four-core and seven-core MCFs prohibit them from being connected directly}, the 7$\times$7 MBS can still be used to close the 4-arm interferometer by using DMUX devices that are compatible with these 7 core fibers as shown in Fig.~\ref{Fig_setup}(c). In this case, the measurement of the  four-dimensional quantum state will have seven different possible outcomes (registered by seven APDs) and, thus, the measurement can be only described as nonprojective. The exact povm that is implemented depends on the unitary matrix representing the 7$\times$7 MBS, as well as which are the four input cores that are connected.  There are a total of 35 inequivalent povms that can be implemented in the configuration given in Fig.~\ref{Fig_setup}(c). The 7$\times$7 MBS used in our experiment was characterized in Ref.~\cite{Optica_2020}, and the corresponding unitary matrix $U_7$ is given explicitly in the \steve{appendix}, along with the 35 possible povms. To implement the protocol described above, and violate the projective bound of Eq.~(\ref{proj}), we optimized the value of $W$ [Eq.~(\ref{W})] over all possible states and projections for each one of the 35 povms we can implement in our scheme. We found that the povm implemented with input cores 4, 5, 6 and 7, namely $M_{4567}$, combined with the states and measurement projections also given explicitly in the \steve{appendix}, allow ones to reach a bound of ${W}_{prtcl}=62.6982$ , which is larger than the projective bound in Eq. (\ref{proj}). 

The proximity of ${W}_{prtcl}$ to $W_{proj}$ requires high-quality phase stabilization and synchronization of preparation and measurement.  To achieve this, the entire system is automatically controlled by two field-programmable gate array electronic units (FPGA1 and FPGA2).  Alice's FPGA1 controls the FMZ used to generate the optical pulses and control her intensity and phase modulators, capable of preparing different predefined states  for each optical pulse sent through the interferometer. Bobs FPGA2 is used to control his intensity/phase modulators and to record the counts of all detectors involved in a given measurement configuration. Independently of the measurement configuration adopted, the four-arm interferometer is distributed in a small area of 30cm$\times$ 30cm with all components thermally insulated to minimize random phase drifts. Nonetheless, phase drifts are always present and a control system is required to actively compensate them. The stabilization system relies on a perturb and observe power point tracking method \cite{Nema} applied to Alice's phase modulators, and is described in more detail in Ref.~\cite{Optica_2020}. Alice's FPGA1 monitors and compensates phase drift of the interferometer every 0.2~s. When a near-zero phase relation is achieved, the system is switched to experiment mode in which the desired states are prepared and measured over 0.1~s. Typically, a total of approximately 10,000 single counts are observed over 0.1s for each measurement configuration.  This procedure allows us to achieve the greater than 99\% interference visibility required for the certification scheme. 

\begin{figure*}[t]
\vspace{0.5cm}
\includegraphics[width=0.99\textwidth]{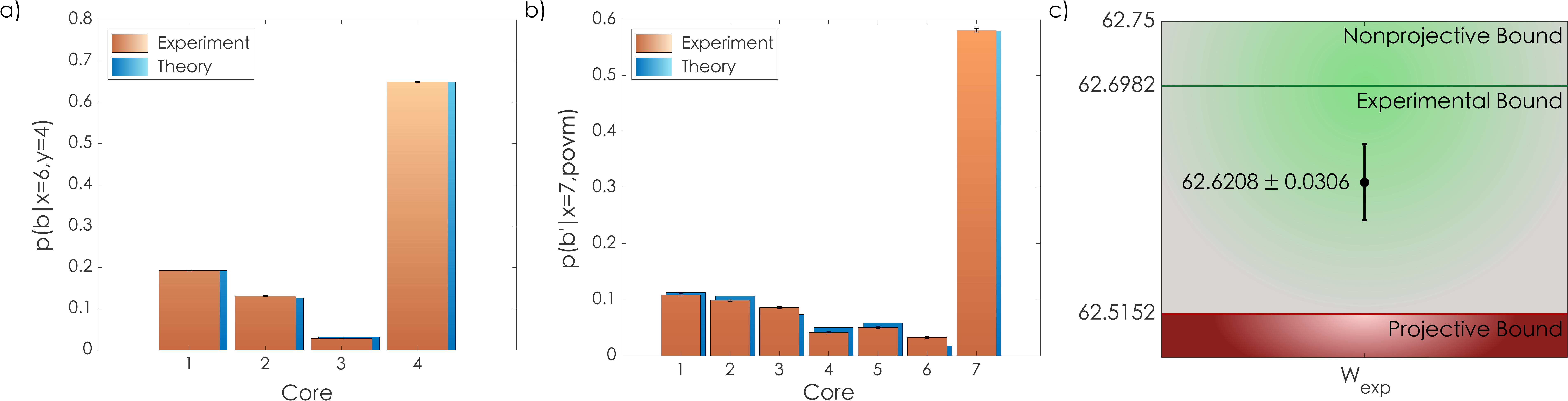}
\caption{Examples of experimental results for certification scheme. a) Measurement probabilities for Bob's projective measurement for inputs $x=6$, $y=4$.  b) Measurement probabilities for Bob's \textbf{povm} measurement for Alice input $x=7$. In both cases, the experimental results are very close to the theoretical predictions. c) Experimentally obtained value of $W$ calculated from (\ref{W}), which clearly lies above the bound achievable with projective measurements. The experimental bound for this protocol is given by $W_{prtcl}$.}
\label{Fig_Results}
\end{figure*}

Figs.~\ref{Fig_Results}(a) and \ref{Fig_Results}(b) show examples of the recorded statistics and compare them with the theoretical predictions for the projective measurement configuration of Fig.~\ref{Fig_setup}(b) and the povm of Fig.~\ref{Fig_setup}(c), respectively. One can clearly see that a high-quality implementation of the protocol is obtained, which is a consequence of the fact that the MCF devices have the same single optical mode as the SMFs, allowing for near perfect mode overlap in all MBSs. All the recorded statistics, related with the many different states and measurements settings required for measuring the value $W$, are given explicitly in the \steve{appendix}.  In Fig.~\ref{Fig_Results}(c) we show the experimental value obtained for $W_{exp}=62.6208\pm0.0036$, which represents a clearly violation of the projective bound with a corresponding $p$-value of $2.793 \times 10^{-4}$, corresponding to 3.45 standard deviations. We can thus certify that generalized quantum measurement in a dimension $d=4$ has been implemented. 
\par

\textit{Conclusion---}
{We have demonstrated that novel multi-core optical fiber devices provide a rich platform for implementing generalized quantum measurements in photonic quantum systems.  In addition to configurability provided by the different input mode configurations possible, the high-quality mode overlap provided by multi-core beam splitters leads to very high-quality measurement fidelity. The latter is key for near-optimal implementations of many quantum information protocols and we demonstrated it in the particularly demanding task of certifying a higher-dimensional generalised quantum measurement in a scenario where only the Hilbert space dimension is known. Our experiment employed a seven-core beam splitter to implement a seven-outcome measurement on a four-dimensional quantum system, achieving measurement fidelity greater than $99.7\%$, certifying that the output results cannot be achieved with projective measurements in a four-dimensional Hilbert space. Our results pave the way towards the realization and use of generalized measurements in higher-dimensional quantum information protocols, a task that has been difficult to achieve until present.}

\begin{acknowledgments}
This work was supported by ANID - Fondo Nacional de Desarrollo Cient\'{i}fico y Tecnol\'{o}gico (FONDECYT) (1180558, 1190901, 1200266, 1200859) and ANID – Millennium Science Initiative Program – ICN17\_012.  JC acknowledges financial support from ANID/REC/PAI77190088.  L.P. was supported by ANID-PFCHA/DOCTORADO-BECAS-CHILE/2019-72200275, the Spanish project PGC2018-094792-B-I00 (MCIU/AEI/FEDER, UE), and CAM/FEDER Project No. S2018/TCS-4342 (QUITEMAD-CM). A.T.~is supported by the Wenner-Gren Foundation.	
 \end{acknowledgments}


%


\appendix
\onecolumngrid

\section{Determining $W_\text{proj}$}
Due to the symmetries of $W$, we can w.~l.~g.~restrict the measurement $\{M_{b'}\}$ to the form $M_{b'}=\ketbra{\phi_{b'}}{\phi_{b'}}$, for $b'\in\{1,2,3,4\}$, where $\{\ket{\phi_{b'}}\}_{b'}$ is an orthonormal basis of $\mathbb{C}^4$, and $M_{b'}=0$ for $b'\in\{5,6,7\}$. The fact that we can restrict to a rank-one projectors for $b'\in\{1,2,3,4\}$ follows from the optimality of Alice preparing pure states $\rho_x=\ketbra{\psi_x}{\psi_x}$ and that the only relevant contribution to the total score is $\bracket{\psi_x}{M_x}{\psi_x}$. Hence, we can write
\begin{equation}
W=\sum_{x,y=1}^7 \left(1+\delta_{x,y}\right)\bracket{\psi_x}{E_{\delta_{x,y}|y}}{\psi_x}+3\sum_{x=1}^4|\braket{\psi_x}{\phi_x}|^2.
\end{equation}
Our optimisation problem becomes
\begin{align}\nonumber
&W_\text{proj}=\max_{\{\psi_x\}, \{E_{b|y}\}, \{\phi_{b'}\}} W.\\\nonumber
& \text{s.t. } \braket{\psi_x}{\psi_x}=1,  \quad \sum_{b'=1}^4 \ketbra{\phi_{b'}}{\phi_{b'}}=\openone\\
& E_{b|y}\geq 0, \qquad E_{0|y}+E_{1|y}=\openone, \qquad E_{b|y}^2=E_{b|y}.
\end{align}
Solving this is difficult. However, increasingly precise upper bounds can be obtained using a hierarchy of semidefinite programming relaxations developed in \cite{Navascues2015}. This method is based on first sampling a basis of moment matrices and then evaluating a semidefinite program over this bases. We refer to \cite{Feix2015} for details.

In our case, we have used a moment matrix of size $176$ which corresponds to the hierarchy level given by
\begin{equation}
\{\openone, \{\psi_x\},\{E_{b|y}\},\{\phi_x\},\{\psi_x E_{b|y}\},\{\psi_x\phi_x\}\}.
\end{equation}
However, this implementation requires over $2000$ sampled moment matrices (at which point the computation was terminated) followed by a correspondingly large semidefinite program. To remedy this, we have employed the method of symmetrisation \cite{Rosset2019}. Specifically, we perform a variable reduction by averaging each sampled moment matrix over the symmetry group $S_4$ of the scoring function, which is simply a permutation of $[1,2,3,4]$ applied to each of $x$, $y$ and $b'$ (when $x,y,b'\in\{5,6,7\}$ we leave them unchanged). We denote by $T_\sigma$ action of a permutation $\sigma\in S_4$ on the moment matrix $\Gamma$. The new moment matrix is then
\begin{equation}
\Gamma\rightarrow \frac{1}{|S_4|}\sum_{\sigma\in S_4} T_\sigma[\Gamma].
\end{equation}
This procedure shrinks the size of the moment matrix basis to just $93$ elements and the computation of the semidefinite program can then be efficiently achieved. The upper bound we obtain is $W_\text{proj}\leq 62.5152$.

To prove that the bound is tight, we use a seesaw approach. That is, we first sample a random set of states $\{\ket{\psi_x}\}$ and evaluate the largest value of $W$ as a semidefinite program over the measurements $\{E_{b|y}\}$  and $\{M_{b'}\}$. Then, for the optimal measurements, we determine the new set of optimal states through an eigenvalue computation. The process is repeated until it appears to converge on a stable value of $W$. The corresponding quantum strategy can then be extracted. Note that since $\{E_{b|y}\}$ is dichotomic and all extremal dichotomic measurements are projective, we need not enforce an additional constraint of projectivity. However, if we impose that $M_{b'}^2=M_{b'}$, then the optimisation over the measurements is no longer a semidefinite program. To circumvent this, we relax the set of projective measurements (recall that we need only to consider the first four outcomes) to the set of all measurements of four outcomes. Running the seesaw procedure we find $W= 62.5152$. We can verify that in the corresponding quantum strategy, $\{M_{b'}\}$ indeed is projective. Thus, we conclude that the bound is tight.

\section{Quantum protocol}\label{AppStrategy}
Here, we outline the states $\rho_x=\ketbra{\psi_x}{\psi_x}$, the dichotomic measurements $\{E_{b|y}\}$ and the seven-outcome measurement $\{M_{b'}\}$ used in the quantum protocol that achieves $W=62.6982$. The protocol is obtained by numerical means. 

The states are given by the following matrix, in which the rows correspond to $\bra{\psi_1},\ldots,\bra{\psi_7}$.
\begin{equation}
\left(
\begin{array}{cccc}
0.507574\, -0.0090103 i & 0.662756\, -0.0213697 i & 0.1439\, -0.0180914 i & 0.530613 \\
-0.0724881+0.498157 i & -0.413325-0.382353 i & 0.260956\, +0.544061 i & 0.255838 \\
-0.415261+0.380216 i & 0.17193\, +0.110383 i & -0.413262-0.139425 i & 0.671584 \\
0.401067\, -0.215166 i & -0.271274+0.36049 i & 0.144303\, +0.431162 i & 0.618532 \\
0.118453\, +0.306095 i & -0.0251893+0.0382293 i & -0.144825-0.846059 i & 0.391649 \\
-0.0518872-0.526847 i & -0.672376+0.320233 i & -0.361049-0.0673529 i & 0.173807 \\
-0.473002-0.355323 i & -0.056111-0.411596 i & 0.211323\, +0.0889052 i & 0.651839 \\
\end{array}
\right)
\end{equation}

The dichotomic measurements are given by the following matrix, in which the rows correspond to the bra-vector onto which $E_{1|y}$ projects, for $y=1,\ldots,7$. Note that $E_{0|y}=\openone-E_{1|y}$.
\begin{equation}
\left(
\begin{array}{cccc}
0.508487\, -0.0150154 i & 0.662197\, -0.0311996 i & 0.151409\, -0.0261662 i & 0.527378 \\
-0.0385109+0.494678 i & -0.443409-0.351319 i & 0.313894\, +0.521516 i & 0.251527 \\
-0.410518+0.378204 i & 0.168252\, +0.119914 i & -0.413796-0.154393 i & 0.671331 \\
0.393413\, -0.230508 i & -0.272517+0.365736 i & 0.137073\, +0.432203 i & 0.615203 \\
0.131541\, +0.30654 i & -0.0324357+0.0247376 i & -0.136997-0.850416 i & 0.380908 \\
-0.00356985-0.523085 i & -0.692675+0.269359 i & -0.367756-0.0818537 i & 0.179085 \\
-0.461476-0.36775 i & -0.0569732-0.4116 i & 0.226916\, +0.0945078 i & 0.647083 \\
\end{array}
\right)
\end{equation}

The generalised measurement employed in the protocol corresponds to operators of the form $M_{b'}=\beta_{b'}\ketbra{\phi_{b'}}{\phi_{b'}}$. The bra-vectors $\bra{\phi_{b'}}$ are the rows of the following matrix, for $b'=1,\ldots,7$
\begin{equation}
\left(
\begin{array}{cccc}
0.505355 & 0.662837 & 0.122035 & 0.538861 \\
0.448439\, +0.262257 i & -0.505467+0.228254 i & 0.589274 & 0.0557004\, -0.268665 i \\
-0.433774+0.384093 i & 0.180038\, +0.0725377 i & -0.407489-0.0925349 i & 0.67233 \\
0.416467\, -0.182098 i & -0.266418+0.346772 i & 0.162878\, +0.429146 i & 0.625676 \\
-0.317427+0.0327698 i & -0.0632815-0.0239012 i & 0.85119 & -0.0741812+0.404431 i \\
-0.165596+0.516406 i & 0.745866 & 0.262609\, +0.230115 i & -0.136976-0.0943827 i \\
-0.499761-0.322502 i & -0.0553898-0.411115 i & 0.176718\, +0.0768536 i & 0.661069 \\
\end{array}
\right).
\end{equation}
The coefficients $\beta_{b'}$ are given by 
\begin{equation}
\begin{pmatrix}
0.5505 & 0.4969 & 0.3651 & 0.6219 & 0.7204 & 0.6638 & 0.5814
\end{pmatrix}
\end{equation}

\section{Measurement Description: the Seven-outcome nonprojective measurement implemented by the $7\times 7$ multiport beam splitter}
\begin{figure}[b]
    \centering
    \includegraphics[width=0.2\textwidth]{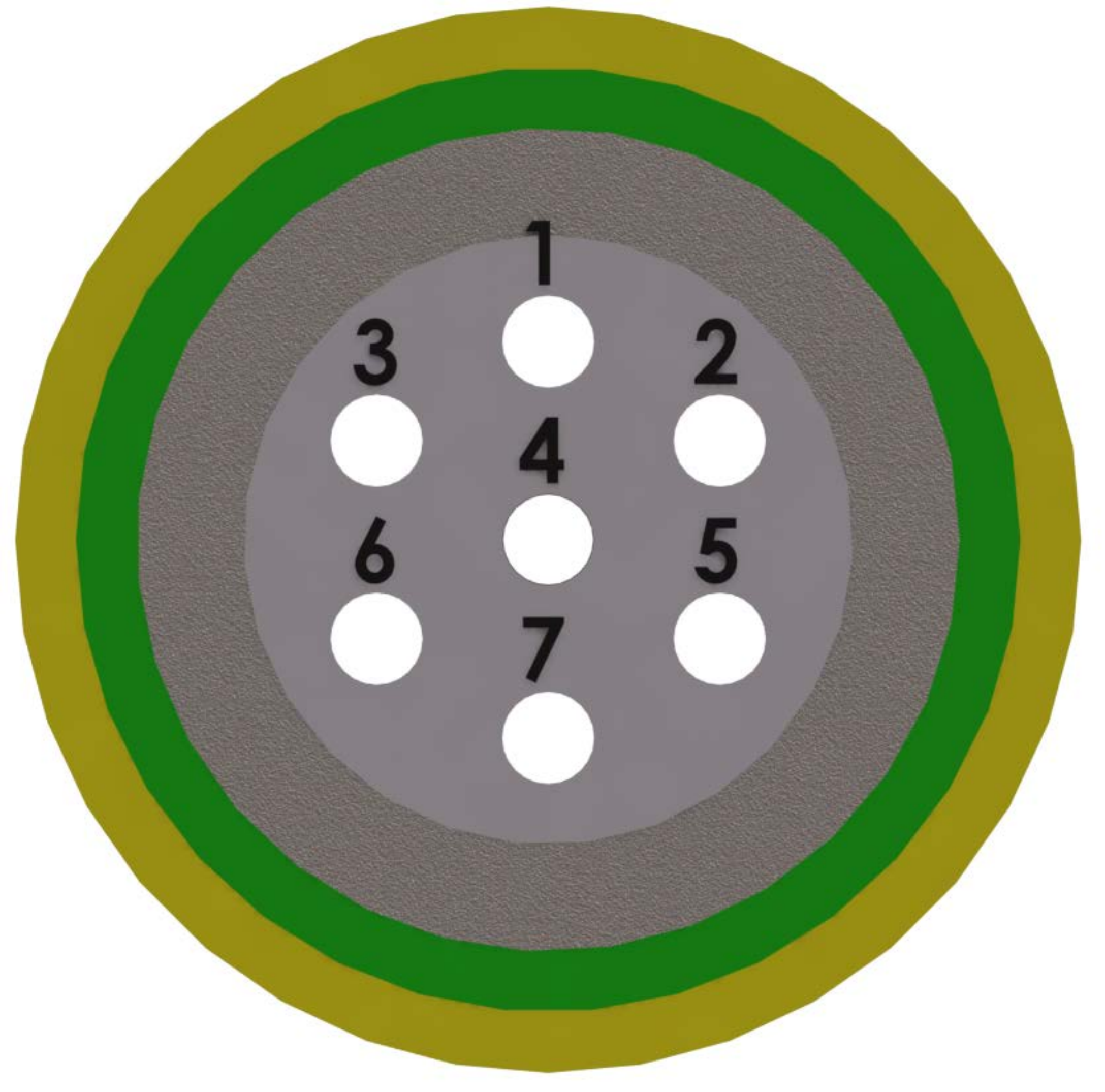}
\caption{A diagram of path-encoding scheme for the multicore fiber in dimension $D=7$. 
\label{figsup:enc}}
\end{figure}

To implement the seven-outcome measurement required to estimate the value of $W$, we resort to using a $7\times 7$ multiport beam splitter (MBS) built-in seven-core multicore fibers (7C-MCFs). We also use a spatial demultiplexer/multiplexer unit (DMUX) for accessing each of the seven cores of the $7\times 7$ MBS independently. These devices allow us to implement the unitary transformation demanded to realize a seven-elements positive operator-valued measurement (POVM). The optical characterization of the $7\times 7$ MBS involves intensity and phase measurements, whose results are numerically optimized to obtain a genuine unitary operation \cite{Optica_2020}. In the logical basis, the unitary matrix $\hat{U}_7$ for the MBS is given by 
{\footnotesize 
\begin{equation}
\hat{U}_7 =
\begin{bmatrix}
0.5639 &   0.2010 &   0.3019 &   0.3749 &   0.4918 &   0.0905 &  0.3998 \\
0.2222 &  -0.0065 + 0.1874i&  -0.5700 - 0.3060i&   0.3558 - 0.0865i&  -0.1447 + 0.3632i&   0.2989 - 0.2884i&  -0.1033 - 0.1635i\\
0.3487 &  -0.6271 - 0.3102i&   0.1178 - 0.0994i&  -0.2245 - 0.2686i&  -0.0469 + 0.1075i&   0.0629 - 0.2445i&  -0.0116 + 0.4061i\\
0.3929 &   0.3320 + 0.0156i&  -0.1620 - 0.2950i&  -0.1267 + 0.3353i&  -0.0489 - 0.3414i&  -0.3319 - 0.1445i&  -0.3447 + 0.3530i\\
0.3709 &  -0.1842 + 0.2868i&  -0.1199 + 0.1069i&  -0.0224 - 0.2699i&   0.0214 - 0.0533i&  -0.0144 + 0.7223i&  -0.3419 - 0.0698i\\
0.1468 &   0.3709 + 0.2029i&   0.3572 - 0.0915i&  -0.0936 - 0.4318i&  -0.5262 + 0.3039i&  -0.2790 - 0.0553i&   0.1351 + 0.0108i\\
0.4444 &  -0.0220 - 0.1704i&  -0.0651 + 0.4328i&  -0.3159 + 0.3254i&  -0.3157 - 0.0201i&   0.0839 - 0.1206i&   0.0989 - 0.4943i
\end{bmatrix}.\nonumber \label{hatU7}
\end{equation}}
Here, $\hat{U}_7$ is written considering the logical basis $\{|j\rangle\}$ as the path-encoding strategy shown in Fig. \ref{figsup:enc}, where $j=1,\ldots,7$ denotes the $j$-th core.

Taking into account the unitary operation $\hat{U}_7$, our MCF-based devices can implement finite sets of inequivalent rank-1 POVM elements, as we explain next. We consider the scenario depicted in Fig. \ref{povmmbs}, where a POVM is realized acting upon an input quantum state $|\Psi\rangle$ of dimension $d$ lower than $D$. In our experiment, $D=7$ corresponds to all the spatial modes available in the $7\times 7$ MBS. Then, the remaining $D-d$ cores define the ancilla system used to extend the Hilbert space dimension of the input state. Likewise, according to Neumark's dilation theorem \cite{NielsenChuang}, the action of the POVM is obtained by performing a projective measurement over the total system (i.e., detecting all cores). The total number of feasible POVMs we can implement is determined by the number of subsets of $d$ elements chosen from a set of $D$ elements, that is
\begin{equation*}
\binom{D}{d}=\frac{D!}{d!(D-d)!}.
\end{equation*}
Then, the rank-1 POVM elements can be written as $\Pi_j=|\chi_j\rangle\langle\chi_j|$, where
\begin{equation*}
|\chi_j\rangle = \hat{M}_{k_1\ldots k_{d}} |j\rangle
\end{equation*}
are unnormalized states. The matrix $\hat{M}_{k_1\ldots k_{d}}$ is the $\hat{U}_7$ restriction on the subspace spanned by the $d$ input modes labeled by $k_1,\ldots k_{d}$. Furthermore, we are able to modify local phases connecting phase modulators at the excited input cores. This operation can be seen as a diagonal matrix $\hat{\Phi}^\dagger_{k_1\ldots k_{d}}$ whose entries represents the phase applied on the input modes, as it also depicted in Fig. \ref{povmmbs}. Phase modulation expands the possibilities to implement even more general POVMs. Nonetheless, no phase modulation at the POVM implementation is required for our task, where we using the stabilization system to ensure this zero-phase condition at the detection stage. We recall that this stabilization scheme actively controls any long term phase drifts that may appear during an experimental run (see the main text for more details).

Then, there are 35 different POVMs that we can implement while four dimensional states ($d=4$) are considered. For the specific task to estimate the value of $W$, it is necessary to find the best POVM to surpass the projective limit $W_{proj}$, we have run optimizations considering these 35 POVMs, founding that the optimal one is 
{\footnotesize \begin{equation}
\hat{M}_{4567}= \begin{bmatrix} 0.3749 & 0.3662 & 0.3501 & 0.3584 & 0.2709 & 0.4418 & 0.4535\\ 0.4918 & -0.2263 - 0.3188i & -0.05239 + 0.1049i & -0.3021 - 0.1664i & 0.05134 - 0.02569i & -0.1856 + 0.5787i & 0.2055 - 0.2405i\\ 0.09054 & 0.3586 + 0.2097i & 0.1473 - 0.2051i & -0.01789 - 0.3615i & -0.7187 + 0.07419i & 0.1132 + 0.261i & -0.145 - 0.02382i\\ 0.3998 & -0.06171 + 0.1833i & -0.3041 + 0.2693i & 0.4521 - 0.1977i & 0.09788 + 0.335i & -0.03915 - 0.1297i & -0.4235 - 0.2733i \end{bmatrix},\nonumber \label{povm4567}
\end{equation}}
which is implemented encoding the input four-dimensional quantum state exciting the cores 4, 5, 6, and 7 at the input of the MBS (see Fig. \ref{figsup:enc}). The full list of the 35 different POVMs is available below.

\section{Measurement Results}
In this section, we show the obtained results for evaluating the optimal value of $W$ in comparison with the expected ones in our experiment. We consider the four-dimensional quantum states prepared by Alice, labeled by the classical input $x\in {1,\ldots,7}$, and the quantum measurements performed by Bob. Here, $y\in {1,\ldots,7}$ denotes a dichotomic projective measurements, and an eighth non projective measurement labeled by $\mathbf{povm}$ corresponds to a seven-outcome measure. Table \ref{table:thproj} shows the 49 expected and recorded probabilities ($p_{th}(b=\delta_{x,y}|x,y)$ and $p_{exp}(b=\delta_{x,y}|x,y)$, respectively) required to evaluate $W$ (see Eq. (1) in the main text) when Bob implements the seven binary projections. 
Likewise, Table \ref{table:povm} shows the relevant expected and recorded probabilities $p(b'=x|x,\mathbf{povm})$ for $W$ when the POVM is implemented. Our experimental results are in good agreement with the expected ones. Moreover, we compute the error values considering Gaussian error propagation and Poissonian statistic for photon counting. Therefore, the experimental value obtained for $W=62.6208\pm0.0306$. 

\begin{table}[!h]
\caption{Expected $p_{th}(b=\delta_{x,y}|x,y)$ and recorded probabilities $p_{exp}(b=\delta_{x,y}|x,y)$ involved in the evaluation of $W$. We consider the case when Alice prepares the $x$ state and Bob implements the $y$ projective measurement which returns the binary outcome $b$.\label{table:thproj}}
\begin{ruledtabular}
\begin{tabular}{cccccccc}
\multirow{2}{*}{} & \multicolumn{7}{c}{$p_{th}(b=\delta_{x,y}|x,y)$} \\
 & $y=1$ & $y=2$ & $y=3$ & $y=4$ & $y=5$ & $y=6$ & $y=7$ \\ \hline
$x=1$ & 0.9998 & 0.9721 & 0.9029 & 0.8411 & 0.9402 & 0.8105 & 0.7870 \\ 
$x=2$ & 0.9730 & 0.9994 & 0.8513 & 0.7764 & 0.8691 & 0.9456 & 0.8118 \\ 
$x=3$ & 0.9098 & 0.8570 & 0.9997 & 0.9840 & 0.7333 & 0.8951 & 0.8292 \\ 
$x=4$ & 0.8370 & 0.7631 & 0.9824 & 0.9996 & 0.9700 & 0.8121 & 0.9053 \\ 
$x=5$ & 0.9377 & 0.8587 & 0.7008 & 0.9695 & 0.9994 & 0.8805 & 0.9151 \\ 
$x=6$ & 0.8014 & 0.9412 & 0.8821 & 0.8079 & 0.8798 & 0.9994 & 0.9753 \\ 
$x=7$ & 0.7907 & 0.8090 & 0.8195 & 0.9094 & 0.9200 & 0.9768 & 0.9995 
\end{tabular}
\bigskip

\begin{tabular}{cccccccc}
\multirow{2}{*}{} & \multicolumn{7}{c}{$p_{exp}(b=\delta_{x,y}|x,y)$} \\
 & $y=1$ & $y=2$ & $y=3$ & $y=4$ & $y=5$ & $y=6$ & $y=7$ \\ \hline
$x=1$ & $0.9937 \pm 0.0008$ & $0.9722 \pm 0.0018$ & $0.9022 \pm 0.0017$ & $0.8409 \pm 0.0016$ & $0.9402 \pm 0.0006$ & $0.8105 \pm 0.0019$ & $0.7870 \pm 0.0019$ \\
$x=2$ & $0.9731 \pm 0.0007$ & $0.9958 \pm 0.0010$ & $0.8523 \pm 0.0040$ & $0.7762 \pm 0.0009$ & $0.8692 \pm 0.0021$ & $0.9454 \pm 0.0012$ & $0.8120 \pm 0.0020$ \\
$x=3$ & $0.9092 \pm 0.0014$ & $0.8569 \pm 0.0009$ & $0.9934 \pm 0.0012$ & $0.9841 \pm 0.0004$ & $0.7345 \pm 0.0048$ & $0.8952 \pm 0.0013$ & $0.8302 \pm 0.0018$ \\
$x=4$ & $0.8372 \pm 0.0021$ & $0.7638 \pm 0.0014$ & $0.9825 \pm 0.0006$ & $0.9970 \pm 0.0009$ & $0.9700 \pm 0.0007$ & $0.8131 \pm 0.0014$ & $0.9058 \pm 0.0011$ \\
$x=5$ & $0.9375 \pm 0.0016$ & $0.8584 \pm 0.0009$ & $0.7013 \pm 0.0007$ & $0.9695 \pm 0.0008$ & $0.9941 \pm 0.0012$ & $0.8806 \pm 0.0023$ & $0.9153 \pm 0.0011$ \\
$x=6$ & $0.8011 \pm 0.0014$ & $0.9414 \pm 0.0009$ & $0.8820 \pm 0.0013$ & $0.8080 \pm 0.0007$ & $0.8803 \pm 0.0023$ & $0.9955 \pm 0.0011$ & $0.9753 \pm 0.0006$ \\
$x=7$ & $0.7899 \pm 0.0018$ & $0.8098 \pm 0.0015$ & $0.8195 \pm 0.0006$ & $0.9098 \pm 0.0007$ & $0.9203 \pm 0.0010$ & $0.9768 \pm 0.0004$ & $0.9951 \pm 0.0012$
\end{tabular}
\end{ruledtabular}
\end{table}

\begin{table}[!h]
\caption{Expected vs the recorded probabilities $p(b'=x|x,\mathbf{povm})$ when the POVM is implemented by Bob obtaining the outcome $b'$, over the state labeled by $x$ sent by Alice.\label{table:povm}}
\begin{ruledtabular}
\begin{tabular}{ccc}
\multirow{2}{*}{$b',x$} & \multicolumn{2}{c}{$p(b'=x|x,\mathbf{povm})$} \\ 
                        & Theory           & Experiment                 \\ \hline
$b',x=1$                & 0.5499           & $0.5495\pm0.0035$          \\
$b',x=2$                & 0.4947           & $0.4912\pm0.0043$          \\
$b',x=3$                & 0.3638           & $0.3637\pm0.0029$          \\
$b',x=4$                & 0.6207           & $0.6206\pm0.0035$          \\
$b',x=5$                & 0.7187           & $0.7152\pm0.0036$          \\
$b',x=6$                & 0.6620           & $0.6623\pm0.0035$          \\
$b',x=7$                & 0.5799           & $0.5814\pm0.0033$         
\end{tabular}
\end{ruledtabular}
\end{table}




\newpage
\section{Full list of the feasible 35 POVMs from the $7\times 7$ MBS}
{\footnotesize\begin{equation*} 
\hat{M}_{1234}= \begin{bmatrix} 0.5639 & 0.2222 & 0.3487 & 0.3929 & 0.3709 & 0.1468 & 0.4444\\ 0.201 & -0.006542 - 0.1874i & -0.6271 + 0.3102i & 0.332 - 0.01557i & -0.1842 - 0.2868i & 0.3709 - 0.2029i & -0.02196 + 0.1704i\\ 0.3019 & -0.57 + 0.306i & 0.1178 + 0.09943i & -0.162 + 0.295i & -0.1199 - 0.1069i & 0.3572 + 0.09154i & -0.0651 - 0.4328i\\ 0.3749 & 0.3558 + 0.08647i & -0.2245 + 0.2686i & -0.1267 - 0.3353i & -0.02242 + 0.2699i & -0.09359 + 0.4318i & -0.3159 - 0.3254i \end{bmatrix}
\end{equation*}} 

{\footnotesize\begin{equation*}  \hat{M}_{1235}= \begin{bmatrix} 0.5639 & 0.2222 & 0.3487 & 0.3929 & 0.3709 & 0.1468 & 0.4444\\ 0.201 & -0.006542 - 0.1874i & -0.6271 + 0.3102i & 0.332 - 0.01557i & -0.1842 - 0.2868i & 0.3709 - 0.2029i & -0.02196 + 0.1704i\\ 0.3019 & -0.57 + 0.306i & 0.1178 + 0.09943i & -0.162 + 0.295i & -0.1199 - 0.1069i & 0.3572 + 0.09154i & -0.0651 - 0.4328i\\ 0.4918 & -0.1447 - 0.3632i & -0.04691 - 0.1075i & -0.04889 + 0.3414i & 0.02136 + 0.0533i & -0.5262 - 0.3039i & -0.3157 + 0.02007i \end{bmatrix}
\end{equation*}} 

{\footnotesize\begin{equation*}  \hat{M}_{1236}= \begin{bmatrix} 0.5639 & 0.2222 & 0.3487 & 0.3929 & 0.3709 & 0.1468 & 0.4444\\ 0.201 & -0.006542 - 0.1874i & -0.6271 + 0.3102i & 0.332 - 0.01557i & -0.1842 - 0.2868i & 0.3709 - 0.2029i & -0.02196 + 0.1704i\\ 0.3019 & -0.57 + 0.306i & 0.1178 + 0.09943i & -0.162 + 0.295i & -0.1199 - 0.1069i & 0.3572 + 0.09154i & -0.0651 - 0.4328i\\ 0.09054 & 0.2989 + 0.2884i & 0.06289 + 0.2445i & -0.3319 + 0.1445i & -0.01444 - 0.7223i & -0.279 + 0.05535i & 0.08389 + 0.1206i \end{bmatrix} 
\end{equation*} }

{\footnotesize\begin{equation*}  \hat{M}_{1237}= \begin{bmatrix} 0.5639 & 0.2222 & 0.3487 & 0.3929 & 0.3709 & 0.1468 & 0.4444\\ 0.201 & -0.006542 - 0.1874i & -0.6271 + 0.3102i & 0.332 - 0.01557i & -0.1842 - 0.2868i & 0.3709 - 0.2029i & -0.02196 + 0.1704i\\ 0.3019 & -0.57 + 0.306i & 0.1178 + 0.09943i & -0.162 + 0.295i & -0.1199 - 0.1069i & 0.3572 + 0.09154i & -0.0651 - 0.4328i\\ 0.3998 & -0.1033 + 0.1635i & -0.01155 - 0.4061i & -0.3447 - 0.353i & -0.3419 + 0.06981i & 0.1351 - 0.01077i & 0.09887 + 0.4943i \end{bmatrix} 
\end{equation*} }

{\footnotesize\begin{equation*}  \hat{M}_{1245}= \begin{bmatrix} 0.5639 & 0.2222 & 0.3487 & 0.3929 & 0.3709 & 0.1468 & 0.4444\\ 0.201 & -0.006542 - 0.1874i & -0.6271 + 0.3102i & 0.332 - 0.01557i & -0.1842 - 0.2868i & 0.3709 - 0.2029i & -0.02196 + 0.1704i\\ 0.3749 & 0.3558 + 0.08647i & -0.2245 + 0.2686i & -0.1267 - 0.3353i & -0.02242 + 0.2699i & -0.09359 + 0.4318i & -0.3159 - 0.3254i\\ 0.4918 & -0.1447 - 0.3632i & -0.04691 - 0.1075i & -0.04889 + 0.3414i & 0.02136 + 0.0533i & -0.5262 - 0.3039i & -0.3157 + 0.02007i \end{bmatrix} 
\end{equation*} }

{\footnotesize\begin{equation*}  \hat{M}_{1246}= \begin{bmatrix} 0.5639 & 0.2222 & 0.3487 & 0.3929 & 0.3709 & 0.1468 & 0.4444\\ 0.201 & -0.006542 - 0.1874i & -0.6271 + 0.3102i & 0.332 - 0.01557i & -0.1842 - 0.2868i & 0.3709 - 0.2029i & -0.02196 + 0.1704i\\ 0.3749 & 0.3558 + 0.08647i & -0.2245 + 0.2686i & -0.1267 - 0.3353i & -0.02242 + 0.2699i & -0.09359 + 0.4318i & -0.3159 - 0.3254i\\ 0.09054 & 0.2989 + 0.2884i & 0.06289 + 0.2445i & -0.3319 + 0.1445i & -0.01444 - 0.7223i & -0.279 + 0.05535i & 0.08389 + 0.1206i \end{bmatrix} 
\end{equation*} }

{\footnotesize\begin{equation*}  \hat{M}_{1247}= \begin{bmatrix} 0.5639 & 0.2222 & 0.3487 & 0.3929 & 0.3709 & 0.1468 & 0.4444\\ 0.201 & -0.006542 - 0.1874i & -0.6271 + 0.3102i & 0.332 - 0.01557i & -0.1842 - 0.2868i & 0.3709 - 0.2029i & -0.02196 + 0.1704i\\ 0.3749 & 0.3558 + 0.08647i & -0.2245 + 0.2686i & -0.1267 - 0.3353i & -0.02242 + 0.2699i & -0.09359 + 0.4318i & -0.3159 - 0.3254i\\ 0.3998 & -0.1033 + 0.1635i & -0.01155 - 0.4061i & -0.3447 - 0.353i & -0.3419 + 0.06981i & 0.1351 - 0.01077i & 0.09887 + 0.4943i \end{bmatrix} 
\end{equation*} }

{\footnotesize\begin{equation*}  \hat{M}_{1256}= \begin{bmatrix} 0.5639 & 0.2222 & 0.3487 & 0.3929 & 0.3709 & 0.1468 & 0.4444\\ 0.201 & -0.006542 - 0.1874i & -0.6271 + 0.3102i & 0.332 - 0.01557i & -0.1842 - 0.2868i & 0.3709 - 0.2029i & -0.02196 + 0.1704i\\ 0.4918 & -0.1447 - 0.3632i & -0.04691 - 0.1075i & -0.04889 + 0.3414i & 0.02136 + 0.0533i & -0.5262 - 0.3039i & -0.3157 + 0.02007i\\ 0.09054 & 0.2989 + 0.2884i & 0.06289 + 0.2445i & -0.3319 + 0.1445i & -0.01444 - 0.7223i & -0.279 + 0.05535i & 0.08389 + 0.1206i \end{bmatrix} 
\end{equation*} }

{\footnotesize\begin{equation*}  \hat{M}_{1257}= \begin{bmatrix} 0.5639 & 0.2222 & 0.3487 & 0.3929 & 0.3709 & 0.1468 & 0.4444\\ 0.201 & -0.006542 - 0.1874i & -0.6271 + 0.3102i & 0.332 - 0.01557i & -0.1842 - 0.2868i & 0.3709 - 0.2029i & -0.02196 + 0.1704i\\ 0.4918 & -0.1447 - 0.3632i & -0.04691 - 0.1075i & -0.04889 + 0.3414i & 0.02136 + 0.0533i & -0.5262 - 0.3039i & -0.3157 + 0.02007i\\ 0.3998 & -0.1033 + 0.1635i & -0.01155 - 0.4061i & -0.3447 - 0.353i & -0.3419 + 0.06981i & 0.1351 - 0.01077i & 0.09887 + 0.4943i \end{bmatrix} 
\end{equation*} }

{\footnotesize\begin{equation*}  \hat{M}_{1267}= \begin{bmatrix} 0.5639 & 0.2222 & 0.3487 & 0.3929 & 0.3709 & 0.1468 & 0.4444\\ 0.201 & -0.006542 - 0.1874i & -0.6271 + 0.3102i & 0.332 - 0.01557i & -0.1842 - 0.2868i & 0.3709 - 0.2029i & -0.02196 + 0.1704i\\ 0.09054 & 0.2989 + 0.2884i & 0.06289 + 0.2445i & -0.3319 + 0.1445i & -0.01444 - 0.7223i & -0.279 + 0.05535i & 0.08389 + 0.1206i\\ 0.3998 & -0.1033 + 0.1635i & -0.01155 - 0.4061i & -0.3447 - 0.353i & -0.3419 + 0.06981i & 0.1351 - 0.01077i & 0.09887 + 0.4943i \end{bmatrix} 
\end{equation*} }

{\footnotesize\begin{equation*}  \hat{M}_{1345}= \begin{bmatrix} 0.5639 & 0.2222 & 0.3487 & 0.3929 & 0.3709 & 0.1468 & 0.4444\\ 0.3019 & -0.57 + 0.306i & 0.1178 + 0.09943i & -0.162 + 0.295i & -0.1199 - 0.1069i & 0.3572 + 0.09154i & -0.0651 - 0.4328i\\ 0.3749 & 0.3558 + 0.08647i & -0.2245 + 0.2686i & -0.1267 - 0.3353i & -0.02242 + 0.2699i & -0.09359 + 0.4318i & -0.3159 - 0.3254i\\ 0.4918 & -0.1447 - 0.3632i & -0.04691 - 0.1075i & -0.04889 + 0.3414i & 0.02136 + 0.0533i & -0.5262 - 0.3039i & -0.3157 + 0.02007i \end{bmatrix} 
\end{equation*} }

{\footnotesize\begin{equation*}  \hat{M}_{1346}= \begin{bmatrix} 0.5639 & 0.2222 & 0.3487 & 0.3929 & 0.3709 & 0.1468 & 0.4444\\ 0.3019 & -0.57 + 0.306i & 0.1178 + 0.09943i & -0.162 + 0.295i & -0.1199 - 0.1069i & 0.3572 + 0.09154i & -0.0651 - 0.4328i\\ 0.3749 & 0.3558 + 0.08647i & -0.2245 + 0.2686i & -0.1267 - 0.3353i & -0.02242 + 0.2699i & -0.09359 + 0.4318i & -0.3159 - 0.3254i\\ 0.09054 & 0.2989 + 0.2884i & 0.06289 + 0.2445i & -0.3319 + 0.1445i & -0.01444 - 0.7223i & -0.279 + 0.05535i & 0.08389 + 0.1206i \end{bmatrix} 
\end{equation*} }

{\footnotesize\begin{equation*}  \hat{M}_{1347}= \begin{bmatrix} 0.5639 & 0.2222 & 0.3487 & 0.3929 & 0.3709 & 0.1468 & 0.4444\\ 0.3019 & -0.57 + 0.306i & 0.1178 + 0.09943i & -0.162 + 0.295i & -0.1199 - 0.1069i & 0.3572 + 0.09154i & -0.0651 - 0.4328i\\ 0.3749 & 0.3558 + 0.08647i & -0.2245 + 0.2686i & -0.1267 - 0.3353i & -0.02242 + 0.2699i & -0.09359 + 0.4318i & -0.3159 - 0.3254i\\ 0.3998 & -0.1033 + 0.1635i & -0.01155 - 0.4061i & -0.3447 - 0.353i & -0.3419 + 0.06981i & 0.1351 - 0.01077i & 0.09887 + 0.4943i \end{bmatrix} 
\end{equation*} }

{\footnotesize\begin{equation*}  \hat{M}_{1356}= \begin{bmatrix} 0.5639 & 0.2222 & 0.3487 & 0.3929 & 0.3709 & 0.1468 & 0.4444\\ 0.3019 & -0.57 + 0.306i & 0.1178 + 0.09943i & -0.162 + 0.295i & -0.1199 - 0.1069i & 0.3572 + 0.09154i & -0.0651 - 0.4328i\\ 0.4918 & -0.1447 - 0.3632i & -0.04691 - 0.1075i & -0.04889 + 0.3414i & 0.02136 + 0.0533i & -0.5262 - 0.3039i & -0.3157 + 0.02007i\\ 0.09054 & 0.2989 + 0.2884i & 0.06289 + 0.2445i & -0.3319 + 0.1445i & -0.01444 - 0.7223i & -0.279 + 0.05535i & 0.08389 + 0.1206i \end{bmatrix} 
\end{equation*} }

{\footnotesize\begin{equation*}  \hat{M}_{1357}= \begin{bmatrix} 0.5639 & 0.2222 & 0.3487 & 0.3929 & 0.3709 & 0.1468 & 0.4444\\ 0.3019 & -0.57 + 0.306i & 0.1178 + 0.09943i & -0.162 + 0.295i & -0.1199 - 0.1069i & 0.3572 + 0.09154i & -0.0651 - 0.4328i\\ 0.4918 & -0.1447 - 0.3632i & -0.04691 - 0.1075i & -0.04889 + 0.3414i & 0.02136 + 0.0533i & -0.5262 - 0.3039i & -0.3157 + 0.02007i\\ 0.3998 & -0.1033 + 0.1635i & -0.01155 - 0.4061i & -0.3447 - 0.353i & -0.3419 + 0.06981i & 0.1351 - 0.01077i & 0.09887 + 0.4943i \end{bmatrix} 
\end{equation*} }

{\footnotesize\begin{equation*}  \hat{M}_{1367}= \begin{bmatrix} 0.5639 & 0.2222 & 0.3487 & 0.3929 & 0.3709 & 0.1468 & 0.4444\\ 0.3019 & -0.57 + 0.306i & 0.1178 + 0.09943i & -0.162 + 0.295i & -0.1199 - 0.1069i & 0.3572 + 0.09154i & -0.0651 - 0.4328i\\ 0.09054 & 0.2989 + 0.2884i & 0.06289 + 0.2445i & -0.3319 + 0.1445i & -0.01444 - 0.7223i & -0.279 + 0.05535i & 0.08389 + 0.1206i\\ 0.3998 & -0.1033 + 0.1635i & -0.01155 - 0.4061i & -0.3447 - 0.353i & -0.3419 + 0.06981i & 0.1351 - 0.01077i & 0.09887 + 0.4943i \end{bmatrix} 
\end{equation*} }

{\footnotesize\begin{equation*}  \hat{M}_{1456}= \begin{bmatrix} 0.5639 & 0.2222 & 0.3487 & 0.3929 & 0.3709 & 0.1468 & 0.4444\\ 0.3749 & 0.3558 + 0.08647i & -0.2245 + 0.2686i & -0.1267 - 0.3353i & -0.02242 + 0.2699i & -0.09359 + 0.4318i & -0.3159 - 0.3254i\\ 0.4918 & -0.1447 - 0.3632i & -0.04691 - 0.1075i & -0.04889 + 0.3414i & 0.02136 + 0.0533i & -0.5262 - 0.3039i & -0.3157 + 0.02007i\\ 0.09054 & 0.2989 + 0.2884i & 0.06289 + 0.2445i & -0.3319 + 0.1445i & -0.01444 - 0.7223i & -0.279 + 0.05535i & 0.08389 + 0.1206i \end{bmatrix} 
\end{equation*} }

{\footnotesize\begin{equation*}  \hat{M}_{1457}= \begin{bmatrix} 0.5639 & 0.2222 & 0.3487 & 0.3929 & 0.3709 & 0.1468 & 0.4444\\ 0.3749 & 0.3558 + 0.08647i & -0.2245 + 0.2686i & -0.1267 - 0.3353i & -0.02242 + 0.2699i & -0.09359 + 0.4318i & -0.3159 - 0.3254i\\ 0.4918 & -0.1447 - 0.3632i & -0.04691 - 0.1075i & -0.04889 + 0.3414i & 0.02136 + 0.0533i & -0.5262 - 0.3039i & -0.3157 + 0.02007i\\ 0.3998 & -0.1033 + 0.1635i & -0.01155 - 0.4061i & -0.3447 - 0.353i & -0.3419 + 0.06981i & 0.1351 - 0.01077i & 0.09887 + 0.4943i \end{bmatrix} 
\end{equation*} }

{\footnotesize\begin{equation*}  \hat{M}_{1467}= \begin{bmatrix} 0.5639 & 0.2222 & 0.3487 & 0.3929 & 0.3709 & 0.1468 & 0.4444\\ 0.3749 & 0.3558 + 0.08647i & -0.2245 + 0.2686i & -0.1267 - 0.3353i & -0.02242 + 0.2699i & -0.09359 + 0.4318i & -0.3159 - 0.3254i\\ 0.09054 & 0.2989 + 0.2884i & 0.06289 + 0.2445i & -0.3319 + 0.1445i & -0.01444 - 0.7223i & -0.279 + 0.05535i & 0.08389 + 0.1206i\\ 0.3998 & -0.1033 + 0.1635i & -0.01155 - 0.4061i & -0.3447 - 0.353i & -0.3419 + 0.06981i & 0.1351 - 0.01077i & 0.09887 + 0.4943i \end{bmatrix} 
\end{equation*} }

{\footnotesize\begin{equation*}  \hat{M}_{1567}= \begin{bmatrix} 0.5639 & 0.2222 & 0.3487 & 0.3929 & 0.3709 & 0.1468 & 0.4444\\ 0.4918 & -0.1447 - 0.3632i & -0.04691 - 0.1075i & -0.04889 + 0.3414i & 0.02136 + 0.0533i & -0.5262 - 0.3039i & -0.3157 + 0.02007i\\ 0.09054 & 0.2989 + 0.2884i & 0.06289 + 0.2445i & -0.3319 + 0.1445i & -0.01444 - 0.7223i & -0.279 + 0.05535i & 0.08389 + 0.1206i\\ 0.3998 & -0.1033 + 0.1635i & -0.01155 - 0.4061i & -0.3447 - 0.353i & -0.3419 + 0.06981i & 0.1351 - 0.01077i & 0.09887 + 0.4943i \end{bmatrix} 
\end{equation*} }

{\footnotesize\begin{equation*}  \hat{M}_{2345}= \begin{bmatrix} 0.201 & 0.1875 & 0.6997 & 0.3324 & 0.3409 & 0.4227 & 0.1718\\ 0.3019 & -0.2859 - 0.5804i & -0.0615 - 0.1414i & -0.1756 + 0.2871i & 0.1548 - 0.04311i & 0.2694 + 0.2517i & -0.4209 + 0.1199i\\ 0.3749 & -0.09883 + 0.3526i & 0.3204 - 0.1412i & -0.1108 - 0.3409i & -0.215 - 0.1647i & -0.2893 + 0.3339i & -0.2824 + 0.3549i\\ 0.4918 & 0.368 - 0.1319i & -0.005624 + 0.1171i & -0.06483 + 0.3387i & -0.05638 - 0.01083i & -0.3158 - 0.5192i & 0.06025 + 0.3105i \end{bmatrix} 
\end{equation*} }

{\footnotesize\begin{equation*}  \hat{M}_{2346}= \begin{bmatrix} 0.201 & 0.1875 & 0.6997 & 0.3324 & 0.3409 & 0.4227 & 0.1718\\ 0.3019 & -0.2859 - 0.5804i & -0.0615 - 0.1414i & -0.1756 + 0.2871i & 0.1548 - 0.04311i & 0.2694 + 0.2517i & -0.4209 + 0.1199i\\ 0.3749 & -0.09883 + 0.3526i & 0.3204 - 0.1412i & -0.1108 - 0.3409i & -0.215 - 0.1647i & -0.2893 + 0.3339i & -0.2824 + 0.3549i\\ 0.09054 & -0.2987 + 0.2887i & 0.05206 - 0.2471i & -0.3383 + 0.1288i & 0.6156 + 0.3782i & -0.2714 - 0.08536i & 0.1089 - 0.09863i \end{bmatrix} 
\end{equation*} }

{\footnotesize\begin{equation*}  \hat{M}_{2347}= \begin{bmatrix} 0.201 & 0.1875 & 0.6997 & 0.3324 & 0.3409 & 0.4227 & 0.1718\\ 0.3019 & -0.2859 - 0.5804i & -0.0615 - 0.1414i & -0.1756 + 0.2871i & 0.1548 - 0.04311i & 0.2694 + 0.2517i & -0.4209 + 0.1199i\\ 0.3749 & -0.09883 + 0.3526i & 0.3204 - 0.1412i & -0.1108 - 0.3409i & -0.215 - 0.1647i & -0.2893 + 0.3339i & -0.2824 + 0.3549i\\ 0.3998 & -0.1598 - 0.1089i & -0.1697 + 0.3691i & -0.3278 - 0.3688i & 0.1261 - 0.3254i & 0.1237 + 0.05538i & 0.4776 - 0.1612i \end{bmatrix} 
\end{equation*} }

{\footnotesize\begin{equation*}  \hat{M}_{2356}= \begin{bmatrix} 0.201 & 0.1875 & 0.6997 & 0.3324 & 0.3409 & 0.4227 & 0.1718\\ 0.3019 & -0.2859 - 0.5804i & -0.0615 - 0.1414i & -0.1756 + 0.2871i & 0.1548 - 0.04311i & 0.2694 + 0.2517i & -0.4209 + 0.1199i\\ 0.4918 & 0.368 - 0.1319i & -0.005624 + 0.1171i & -0.06483 + 0.3387i & -0.05638 - 0.01083i & -0.3158 - 0.5192i & 0.06025 + 0.3105i\\ 0.09054 & -0.2987 + 0.2887i & 0.05206 - 0.2471i & -0.3383 + 0.1288i & 0.6156 + 0.3782i & -0.2714 - 0.08536i & 0.1089 - 0.09863i \end{bmatrix} 
\end{equation*} }

{\footnotesize\begin{equation*}  \hat{M}_{2357}= \begin{bmatrix} 0.201 & 0.1875 & 0.6997 & 0.3324 & 0.3409 & 0.4227 & 0.1718\\ 0.3019 & -0.2859 - 0.5804i & -0.0615 - 0.1414i & -0.1756 + 0.2871i & 0.1548 - 0.04311i & 0.2694 + 0.2517i & -0.4209 + 0.1199i\\ 0.4918 & 0.368 - 0.1319i & -0.005624 + 0.1171i & -0.06483 + 0.3387i & -0.05638 - 0.01083i & -0.3158 - 0.5192i & 0.06025 + 0.3105i\\ 0.3998 & -0.1598 - 0.1089i & -0.1697 + 0.3691i & -0.3278 - 0.3688i & 0.1261 - 0.3254i & 0.1237 + 0.05538i & 0.4776 - 0.1612i \end{bmatrix} 
\end{equation*} }

{\footnotesize\begin{equation*}  \hat{M}_{2367}= \begin{bmatrix} 0.201 & 0.1875 & 0.6997 & 0.3324 & 0.3409 & 0.4227 & 0.1718\\ 0.3019 & -0.2859 - 0.5804i & -0.0615 - 0.1414i & -0.1756 + 0.2871i & 0.1548 - 0.04311i & 0.2694 + 0.2517i & -0.4209 + 0.1199i\\ 0.09054 & -0.2987 + 0.2887i & 0.05206 - 0.2471i & -0.3383 + 0.1288i & 0.6156 + 0.3782i & -0.2714 - 0.08536i & 0.1089 - 0.09863i\\ 0.3998 & -0.1598 - 0.1089i & -0.1697 + 0.3691i & -0.3278 - 0.3688i & 0.1261 - 0.3254i & 0.1237 + 0.05538i & 0.4776 - 0.1612i \end{bmatrix} 
\end{equation*} }

{\footnotesize\begin{equation*}  \hat{M}_{2456}= \begin{bmatrix} 0.201 & 0.1875 & 0.6997 & 0.3324 & 0.3409 & 0.4227 & 0.1718\\ 0.3749 & -0.09883 + 0.3526i & 0.3204 - 0.1412i & -0.1108 - 0.3409i & -0.215 - 0.1647i & -0.2893 + 0.3339i & -0.2824 + 0.3549i\\ 0.4918 & 0.368 - 0.1319i & -0.005624 + 0.1171i & -0.06483 + 0.3387i & -0.05638 - 0.01083i & -0.3158 - 0.5192i & 0.06025 + 0.3105i\\ 0.09054 & -0.2987 + 0.2887i & 0.05206 - 0.2471i & -0.3383 + 0.1288i & 0.6156 + 0.3782i & -0.2714 - 0.08536i & 0.1089 - 0.09863i \end{bmatrix} 
\end{equation*} }

{\footnotesize\begin{equation*}  \hat{M}_{2457}= \begin{bmatrix} 0.201 & 0.1875 & 0.6997 & 0.3324 & 0.3409 & 0.4227 & 0.1718\\ 0.3749 & -0.09883 + 0.3526i & 0.3204 - 0.1412i & -0.1108 - 0.3409i & -0.215 - 0.1647i & -0.2893 + 0.3339i & -0.2824 + 0.3549i\\ 0.4918 & 0.368 - 0.1319i & -0.005624 + 0.1171i & -0.06483 + 0.3387i & -0.05638 - 0.01083i & -0.3158 - 0.5192i & 0.06025 + 0.3105i\\ 0.3998 & -0.1598 - 0.1089i & -0.1697 + 0.3691i & -0.3278 - 0.3688i & 0.1261 - 0.3254i & 0.1237 + 0.05538i & 0.4776 - 0.1612i \end{bmatrix} 
\end{equation*} }

{\footnotesize\begin{equation*}  \hat{M}_{2467}= \begin{bmatrix} 0.201 & 0.1875 & 0.6997 & 0.3324 & 0.3409 & 0.4227 & 0.1718\\ 0.3749 & -0.09883 + 0.3526i & 0.3204 - 0.1412i & -0.1108 - 0.3409i & -0.215 - 0.1647i & -0.2893 + 0.3339i & -0.2824 + 0.3549i\\ 0.09054 & -0.2987 + 0.2887i & 0.05206 - 0.2471i & -0.3383 + 0.1288i & 0.6156 + 0.3782i & -0.2714 - 0.08536i & 0.1089 - 0.09863i\\ 0.3998 & -0.1598 - 0.1089i & -0.1697 + 0.3691i & -0.3278 - 0.3688i & 0.1261 - 0.3254i & 0.1237 + 0.05538i & 0.4776 - 0.1612i \end{bmatrix} 
\end{equation*} }

{\footnotesize\begin{equation*}  \hat{M}_{2567}= \begin{bmatrix} 0.201 & 0.1875 & 0.6997 & 0.3324 & 0.3409 & 0.4227 & 0.1718\\ 0.4918 & 0.368 - 0.1319i & -0.005624 + 0.1171i & -0.06483 + 0.3387i & -0.05638 - 0.01083i & -0.3158 - 0.5192i & 0.06025 + 0.3105i\\ 0.09054 & -0.2987 + 0.2887i & 0.05206 - 0.2471i & -0.3383 + 0.1288i & 0.6156 + 0.3782i & -0.2714 - 0.08536i & 0.1089 - 0.09863i\\ 0.3998 & -0.1598 - 0.1089i & -0.1697 + 0.3691i & -0.3278 - 0.3688i & 0.1261 - 0.3254i & 0.1237 + 0.05538i & 0.4776 - 0.1612i \end{bmatrix} 
\end{equation*} }

{\footnotesize\begin{equation*}  \hat{M}_{3456}= \begin{bmatrix} 0.3019 & 0.647 & 0.1542 & 0.3365 & 0.1607 & 0.3687 & 0.4377\\ 0.3749 & -0.2726 - 0.2445i & 0.001653 + 0.3501i & -0.2329 + 0.2725i & -0.1629 - 0.2164i & 0.01653 + 0.4415i & 0.3688 - 0.264i\\ 0.4918 & -0.0443 + 0.3884i & -0.1052 - 0.05189i & 0.3228 - 0.1215i & -0.05141 - 0.02556i & -0.5852 - 0.1638i & 0.02711 - 0.3152i\\ 0.09054 & -0.1269 - 0.3955i & 0.2058 + 0.1463i & 0.2864 + 0.2213i & 0.4915 + 0.5295i & -0.2566 + 0.1229i & -0.1318 + 0.06502i \end{bmatrix} 
\end{equation*} }

{\footnotesize\begin{equation*}  \hat{M}_{3457}= \begin{bmatrix} 0.3019 & 0.647 & 0.1542 & 0.3365 & 0.1607 & 0.3687 & 0.4377\\ 0.3749 & -0.2726 - 0.2445i & 0.001653 + 0.3501i & -0.2329 + 0.2725i & -0.1629 - 0.2164i & 0.01653 + 0.4415i & 0.3688 - 0.264i\\ 0.4918 & -0.0443 + 0.3884i & -0.1052 - 0.05189i & 0.3228 - 0.1215i & -0.05141 - 0.02556i & -0.5852 - 0.1638i & 0.02711 - 0.3152i\\ 0.3998 & 0.1683 - 0.09526i & -0.2707 - 0.3029i & -0.1435 + 0.4721i & 0.2088 - 0.2797i & 0.1282 - 0.04397i & -0.5035 + 0.02425i \end{bmatrix} 
\end{equation*} }

{\footnotesize\begin{equation*}  \hat{M}_{3467}= \begin{bmatrix} 0.3019 & 0.647 & 0.1542 & 0.3365 & 0.1607 & 0.3687 & 0.4377\\ 0.3749 & -0.2726 - 0.2445i & 0.001653 + 0.3501i & -0.2329 + 0.2725i & -0.1629 - 0.2164i & 0.01653 + 0.4415i & 0.3688 - 0.264i\\ 0.09054 & -0.1269 - 0.3955i & 0.2058 + 0.1463i & 0.2864 + 0.2213i & 0.4915 + 0.5295i & -0.2566 + 0.1229i & -0.1318 + 0.06502i\\ 0.3998 & 0.1683 - 0.09526i & -0.2707 - 0.3029i & -0.1435 + 0.4721i & 0.2088 - 0.2797i & 0.1282 - 0.04397i & -0.5035 + 0.02425i \end{bmatrix} 
\end{equation*} }

{\footnotesize\begin{equation*}  \hat{M}_{3567}= \begin{bmatrix} 0.3019 & 0.647 & 0.1542 & 0.3365 & 0.1607 & 0.3687 & 0.4377\\ 0.4918 & -0.0443 + 0.3884i & -0.1052 - 0.05189i & 0.3228 - 0.1215i & -0.05141 - 0.02556i & -0.5852 - 0.1638i & 0.02711 - 0.3152i\\ 0.09054 & -0.1269 - 0.3955i & 0.2058 + 0.1463i & 0.2864 + 0.2213i & 0.4915 + 0.5295i & -0.2566 + 0.1229i & -0.1318 + 0.06502i\\ 0.3998 & 0.1683 - 0.09526i & -0.2707 - 0.3029i & -0.1435 + 0.4721i & 0.2088 - 0.2797i & 0.1282 - 0.04397i & -0.5035 + 0.02425i \end{bmatrix} 
\end{equation*} }

{\footnotesize\begin{equation*}  \hat{M}_{4567}= \begin{bmatrix} 0.3749 & 0.3662 & 0.3501 & 0.3584 & 0.2709 & 0.4418 & 0.4535\\ 0.4918 & -0.2263 - 0.3188i & -0.05239 + 0.1049i & -0.3021 - 0.1664i & 0.05134 - 0.02569i & -0.1856 + 0.5787i & 0.2055 - 0.2405i\\ 0.09054 & 0.3586 + 0.2097i & 0.1473 - 0.2051i & -0.01789 - 0.3615i & -0.7187 + 0.07419i & 0.1132 + 0.261i & -0.145 - 0.02382i\\ 0.3998 & -0.06171 + 0.1833i & -0.3041 + 0.2693i & 0.4521 - 0.1977i & 0.09788 + 0.335i & -0.03915 - 0.1297i & -0.4235 - 0.2733i \end{bmatrix} 
\end{equation*} }

\end{document}